\documentclass[amsmath,amssymb,twocolumn,nofootinbib]{revtex4-1}
\usepackage{graphicx}
\usepackage{dcolumn}
\usepackage{bm}
\usepackage{amsmath}
\usepackage{amssymb}
\usepackage[utf8]{inputenc}
\usepackage[T1]{fontenc}
\usepackage{mathptmx}
\usepackage{etoolbox}
\usepackage{tikz}

\usepackage{fancyhdr}
\usepackage{fancybox}
\usepackage{float}
\usepackage{mathtools,amssymb,amsthm,amsmath,bm}
\usepackage{array}
\usepackage{xcolor}
\usepackage{tabularx}
\usepackage{multirow}
\usepackage{geometry}
\usepackage{hyperref}
\usepackage{booktabs,tabularx,graphicx}
\newcolumntype{C}{>{\centering\arraybackslash}X}
\newcommand{\Eq}[1]{Eq.~\ref{eq:#1}}

\newcommand{\Tab}[1]{Table~\ref{tab:#1}}

\newcommand{\Mc}{\mathcal}

\renewcommand{\div}[1]{\nabla \cdot \left(#1\right)}
\newcommand{\Lapl}[1]{\Delta #1}

\newcommand{\change}[1]{{#1}}

\newcounter{rownumbers}

\makeatletter
\def\@email#1#2{%
 \endgroup
 \patchcmd{\titleblock@produce}
  {\frontmatter@RRAPformat}
  {\frontmatter@RRAPformat{\produce@RRAP{*#1\href{mailto:#2}{#2}}}\frontmatter@RRAPformat}
}
\makeatother
\begin{document}

\title{Rise and fall of a multicomponent droplet in a surrounding fluid: simulation study of a bumpy path}
\author{Mirantsoa Aimé Rasolofomanana}
\affiliation{CEA, DES, IRESNE, DTN, St Paul-Lez-Durance 13108, France}
\affiliation{Laboratoire de Physique de la Matière Condensée, Institut Polytechnique de Paris, CNRS, Palaiseau 91120, France}

\author{Romain Le Tellier}%
\affiliation{CEA, DES, IRESNE, DTN, St Paul-Lez-Durance 13108, France}


\author{Hervé Henry}
\affiliation{Laboratoire de Physique de la Matière Condensée, Institut Polytechnique de Paris, CNRS,  Palaiseau 91120, France}

\date{\today}

\begin{abstract}
The coupling between mass transfer and hydrodynamic phenomena in two-phase flow is not straightforward due to the different effects that can be encountered. Its description is complex and requires particular efforts, especially in the modelling of the interface between phases. Here, we consider the case of a two components droplet   (one miscible and one immiscible in water) released in a 2D rectangular domain filled with water. Mass transfer occurs between the miscible element and the surrounding water, which leads to a density inversion that directly affects the droplet trajectory through buoyancy. We perform simulations using a ternary Cahn-Hilliard model to capture such coupled phenomena. The Boussinesq approximation for a multicomponent system is used to define the density law and an analytical chemical potential is proposed for the thermodynamic landscape. The effect of the mobility parameter on the flow is highlighted, the solutal Marangoni effect is discussed and the results are found in good agreement with the dynamics described from an experimental study of the literature. 
\end{abstract}
\maketitle

\section{Introduction}\label{Introduction}
 Multiphase flows are ubiquitous and can be seen in industrial processes and natural environment. They are difficult to describe and model numerically due to the coupling between the moving free surface and the fluid flow. This leads to complex behaviours that are still a challenge to describe and they are therefore driving new development in modelling. This complexity is augmented when mass transfer between the two phases  occurs. This is typically the case in industrial flows and mixing. For instance, this takes place in the pool of molten materials (corium) formed during a severe nuclear accident and affect thermal transfers \cite{zanella2021,asmolov2003,barrachin2004}. It can also occur in microfluidic devices \cite{Grossier} and also during the geological sequestration of CO$_2$ either in coal or in deep aquifers \cite{CO2a,CO2b}. Finally, they are of importance in industrial processes where the dispersion of chemical components in a surrounding fluid is mediated by the motion of droplets \cite{wegener2014}.

 In such a situation, mass transfer will occur between the two phases and will affect the fluid flow. This was exemplified in a recent experimental work \cite{rao2015} inspired by oil spills. Here, through numerical simulations, we show the ability of a ternary phase field model to describe the complex interplay of mass transfer through diffusion and flow equation in the presence of buoyant force. To this purpose, we compare numerical results obtained with a recent experimental work in a qualitative manner. The numerical results also show that the difference between a \textit{naive} description of the motion and the fully coupled problem can be dramatic and can be well explained by the analysis of the solution of the fully coupled problem. This illustrates the need for fully coupled simulations of such systems. Indeed,  simulations methods such as volume of fluids \cite{VOF0} or level set usually rely on the assumption that both phases are immiscible and that no mass transfer occurs through the interface \cite{Multireview}. Phase field models based on either the Allen-Cahn \cite{anderson2000,folch2005} or the Cahn-Hilliard model \cite{cahn1958,kim2005d,toth2016} can describe mass transfer through the interface. However, this possibility has not been used fully explored and is seen as an adverse effect when modelling multiphase flows of perfectly immiscible fluids \cite{magaletti2013,zanella2020}. Here, we show that a ternary Cahn-Hilliard model coupled with the Navier-Stokes equations, using a simple free energy functional is able to describe the complex interplay between fluid flow and diffusive transport and that this modelling effort brings new insights. This is exemplified by numerical simulations of the motion of a multicomponent droplet moving into water while one of its components diffuses out of it leading to changes in the droplet mass density. The numerical results are compared to experimental results presented in \cite{rao2015}.

The paper is organised as follows. First, we briefly describe the phenomena at play in the experiment of interest and discuss qualitatively why a proper description of the coupling between mass transfer and hydrodynamics is needed. Then, we describe the generic form of the model and the simulation setup. We  discuss the choice of parameters and propose a convergence study with respect to the interface thickness. Thereafter, we present numerical simulation results that show the ability of the model to describe the interplay between hydrodynamics and mass transfer phenomena before drawing a conclusion and perspectives.

\section{Description of the physical problem}

The dynamics of a multicomponent droplet released in a quiescent water column is studied experimentally in \cite{rao2015}. The droplet is initially composed of two components: one (denoted A) with a density lower than of water is miscible with water, the other (denoted B) with a density larger than of water is not. For some compositions of the droplet, the following sequence is observed: first, the droplet rises in the water and slows down until it reaches zero velocity and finally sinks with an increasing velocity until a limiting velocity is reached. This can be well explained qualitatively by the evolution of the composition of the droplet due to diffusion.

Indeed, if the droplet composition is sufficiently rich in the light element, it is lighter than water and  will move up.  Because of the  diffusion of the miscible element in water, the amount of A in the droplet  decreases while the amount of heavy element B remains constant. As a result, the density of the droplet  increases until it reaches the density of pure B which is heavier than water. This implies that the density of the droplet that was initially less than water increases, becomes equal to that of the surrounding water and eventually is larger, resulting in the observed motion.

This explanation is fairly simple, however, if one wants to reproduce this behaviour quantitatively, which may be of interest for instance when modelling flow patterns in corium \cite{zanella2021}, the crude approximation where the diffusive mass transfer is computed considering a droplet in pure water and then the motion of the droplet is obviously incorrect, since it neglects the diffusion layer that builds around the droplet. In the same spirit, considering that the mass transfer is the same as for a quiescent droplet is incorrect, since it neglects the advection that affects the composition around the droplet. Therefore, in order to reproduce quantitatively such motion, one needs to solve the fully coupled problem: the advection-diffusion equation together with the Navier-Stokes equations. This can be done with numerical simulations using the phase field method \cite{kim2005d,toth2016,lee2021,vasilopoulos2020}. This class of models relies on a diffuse interface that can be tracked implicitly and has been used to model multiphase flows but rarely to model diffusive transport coupled with flow while it has been shown to be efficient \cite{tree2017}.  

\section{Description of the model}

The phase field model  is a diffuse interface approach. It presents the advantage of being thermodynamically consistent: by construction, it derives from a free energy functional (the thermodynamic landscape). In addition, it is  simple to implement since it is based on coupled PDEs. Nevertheless, the thermodynamic landscape and model parameters must be chosen with care to give the right properties. For instance, the diffuse interface thickness is given by a balance between the thermodynamic landscape and the magnitude of a squared gradient term. The accuracy of the model when considering simple multiphase flows has been shown to depend on the diffuse interface thickness and on the mobility of chemical species \cite{magaletti2013,zanella2020}. In the more complex case of ternary system, additional interface properties, such as adsorption, have been shown to be difficult to control for thermodynamic landscapes obtained from thermodynamic databases such as the ones constructed by the popular Calphad method \cite{cardon2016,rasolofomanana2022a}. 
 
Here, we use the phase field model to study numerically the dynamics of a multicomponent droplet in water. In such approaches, material properties vary continuously from one phase to the other and the interface is defined implicitly as the isosurface of an order parameter. The diffuse interface description was originally proposed by Van Der Waals \cite{rowlinson1979}, it was then popularised by Cahn \cite{cahn1958} in the context of diffusive mass transfer with a conserved order parameter. It has been then used, with a non conserved order parameter, to describe the motion of free interfaces in a natural way \cite{Levine1986,Caginalp} in the context of solid growth where it has become a standard approach \cite{wu2022,kadambi2020}. It has been extended thereafter to fluid flow using either a conserved order parameter \cite{anderson2000} or a non-conserved order parameter \cite{folch2005,Folch1999}. Since the diffusion of chemical species in a ternary system is considered, we use a conserved approach that allows a simple use of ternary system \cite{toth2016} and  that has already been used in other contexts \cite{tree2017}. It must be noted that in this conserved approach  the anticurvature term introduced for non conserved models in \cite{Folch1999} is not needed. In an effort to describe actual systems, there has been recent efforts to couple the principle of such diffuse interface model with free energy functional extracted from thermodynamic databases \cite{cardon2016,rasolofomanana2022a}. In this paper, we focus on the description of the interplay between diffusive mass transport and fluid flow. Therefore, we have chosen a free energy functional that is prone to very little interface adsorption. 

\begin{figure}
	\centering
		\includegraphics[width=0.48\textwidth]{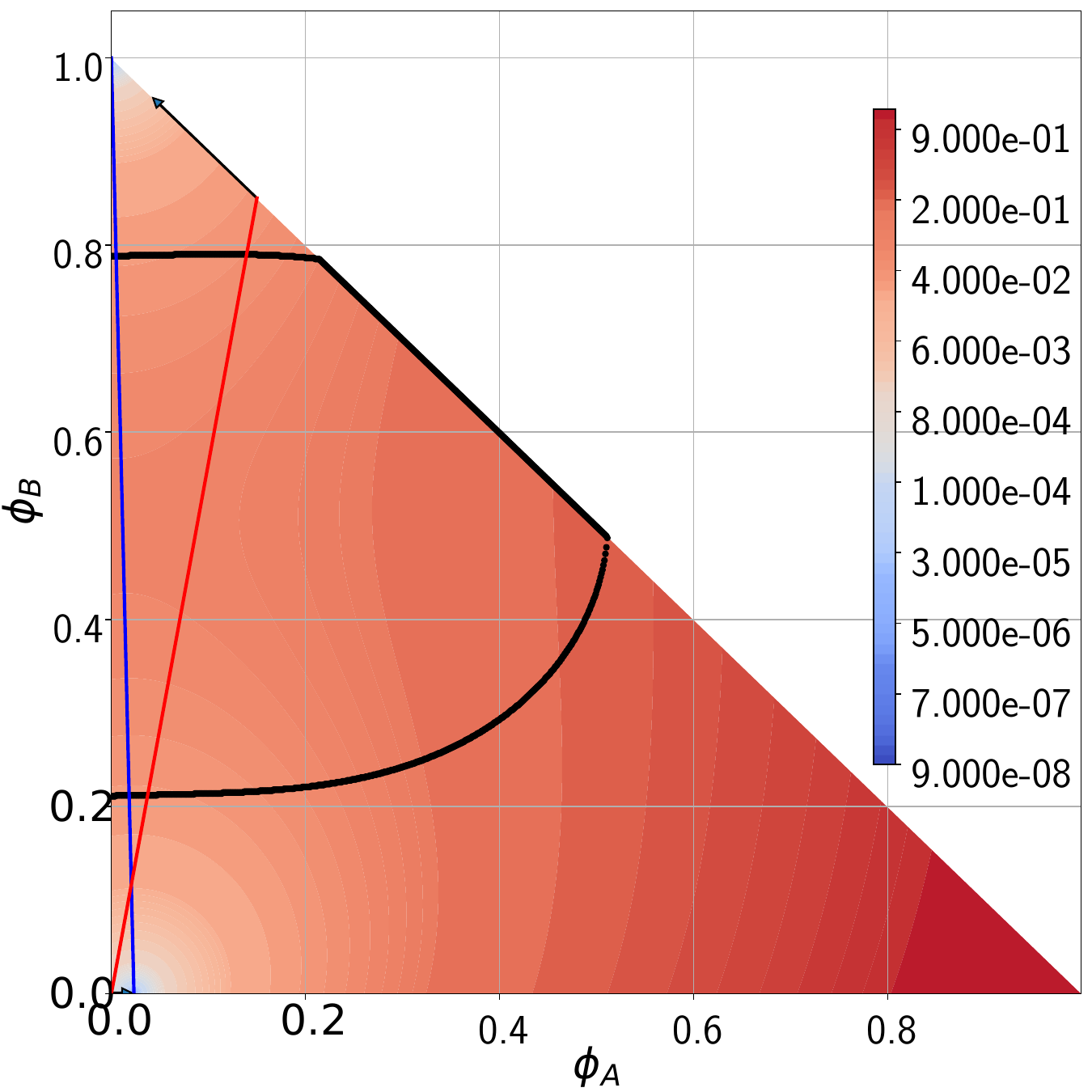} 
	\caption{Thermodynamic landscape represented in the energy surface of the free energy density $G$ shown on logarithmic scale. The blue line is the tie-line, the red one connects the compositions at initial state and the black line represents the spinodal curve.}\label{fig:paysage thermodynamique0.15_0.85}
\end{figure}

In the model, three order parameters $\phi_{i\in A,\ B,\ \mbox{water}}$ represent the component volume fractions of element A, element B and water. The molar volume is assumed constant for each element and independent of the composition. The evolution of these conserved order parameters  $\phi_{i}$ for i$\in$\{A,B\} are governed by the advective Cahn-Hilliard equations:
\begin{eqnarray}
\frac{\partial \phi_{i}}{\partial{t}} + \bm{u}\cdot\nabla\phi_i &=& \div{\sum_{j \in \left\{A,B \right\}}\Mc{M}_{i,j} \nabla\tilde{\mu}_{j}} \nonumber \\ \label{eq:CH1} 
\tilde{\mu}_{i} &=& \lambda\frac{\partial {G}}{\partial \phi_i}-\sum_{j \in \left\{A,B \right\}}\kappa_{i,j}\Lapl{\phi_j}
 \label{eq:CH2}
\end{eqnarray}	
with $\bm{u}$ the advective velocity field, $\tilde{\mu}_j$ the chemical potential of the element $j$, ${G}$ the double-well free energy, the mobility $\mathcal{M}_{i,j}$, the gradient energy $\kappa_{i,j}$ and $\lambda$ an upscaling parameter \cite{cardon2016,rasolofomanana2022a} defined later. The gradient coefficient matrix is chosen simply to be
\begin{equation}
\bar{\bar{\kappa}} = 
\begin{bmatrix}
\kappa^{bin} & 0 \\
0 & \kappa^{bin}
\end{bmatrix}
\end{equation} 
where $\kappa^{bin}$ is the gradient energy coefficient obtained from a binary system and defined explicitly later. The coefficients of the mobility matrix are taken simply to be equal to:
\begin{equation}
\bar{\bar{\mathcal{M}}} = 
\begin{bmatrix}
M & 0 \\
0 & M
\end{bmatrix}
\label{eq:mobility}
\end{equation} 
The choice of $M$, for a given set of parameters is related to the diffusion constant in the bulk phases (either the droplet or the water phase) and is discussed in section \ref{secparamconv}. Here we have chosen to refer to $D_A$ the diffusion constant of the miscible element in water. 

These advective Cahn-Hilliard equations are coupled with the Navier-Stokes equations in order to describe the whole system. Based on the so-called ``one fluid model'', the phase field model treats the two phases as a single one \cite{mirjalili2017}. Thus, we get the Navier-Stokes equations for a ternary system under the Boussinesq approximation:
\begin{widetext}
\begin{equation}
\rho^{\star}\frac{D \bm{u}}{Dt}  =  -\nabla P + \sum_{i \in \left\{A,B \right\}}\tilde{\mu}_i\nabla \phi_{i}+ \displaystyle\Big(\rho(\phi_{A},\phi_{B})-\rho^{\star}\Big)\bm{g} + \eta \Lapl{\bm{u}} \label{eq:NS1}
\end{equation}
\end{widetext}
where $D/Dt = \frac{\partial}{\partial t} + \bm{u} \nabla \cdot$, $P$ is the pressure, $\eta$ is the dynamic viscosity (taken independent of the chemical composition here), $\bm{g}$ is the gravity, the term with $\tilde{\mu}_i$ represents the capillary force within the interface as proposed in \cite{kim2012a}. Once integrated through the  interface at equilibrium, the capillary force leads to the surface tension. $\rho^{\star}$ is the mass density of reference and $\rho$ is the mass density that is function of the order parameters $\phi_{A}$ and $\phi_{B}$ and writes, under the assumption of constant molar volume:
\begin{eqnarray}
\rho(\phi_{A},\phi_{B})& = &\rho^{\star}(1+\displaystyle\sum_{i\in\{A,B\}}\beta_i\phi_{i})\nonumber\\ 
&=& \rho^{\star}\left(1+\beta_{A} \phi_{A} + \beta_{B} \phi_{B} \right)
\label{eq:Boussinesq_approx}
\end{eqnarray}
with $ \rho^{\star}=$1000 kg/m$^3$, $\beta_A=-0.416$ and $\beta_B=0.05$, so that the density of pure B is 1050 kg.m$^{-3}$ and the density of pure A is 584 kg.m$^{-3}$.

The chemical potential (\Eq{CH2}) is given by the double-well free energy that is chosen in such a way that at thermodynamic equilibrium, the droplet contains only pure B element and the element A is completely dissolved in the continuous phase. It writes:
\begin{equation}
{G}=\left[\left(\phi_A-\phi_A^{eq,cont}\right)^{2}+\phi_B^2 \right]\left[\left(1-\phi_B\right)^2+ \phi_A^2 \right]
\label{eq:free_energy}
\end{equation}
 where $\phi_A^{eq,cont}$ is the molar fraction of the element A in the continuous phase at thermodynamic equilibrium. With such a choice of thermodynamic landscape, there exists a region where the homogeneous mixture is unstable (the spinodal domain). It corresponds to the domain for which the Hessian matrix of $G$ has one negative eigenvalue and is represented in fig. \ref{fig:paysage thermodynamique0.15_0.85}. In this region, small fluctuations of the composition can grow and lead to the formation of domains with different compositions. It should be noted that with this choice of ${G}$, no adsorption at the interface between pure B and pure water occurs when using a diagonal $\bar{\bar{\kappa}}$. 

\begin{table}
  $$
  \begin{array}{|c|c||c|c|}
    \hline
    \eta & 10^-3\ \mbox{Pa.s}& \sigma& 0.0036\ \mbox{J.m}^{-2}\\
    \epsilon & 1\mbox{ to }8\  10^{-4}\mbox{m} & D_A& 5.4\,10^{-7}\mbox{ to } 1.08\,10{-5} \mbox{ m}^2.\mbox{s}^{-1}  \\ 
    \beta_A& -0.416 & \beta_B & 0.05 \\
    \rho_A & 584\  \mbox{kg.m}^{-3}& \rho_B &1050\  \mbox{kg.m}^{-3}\\
    \rho^{\star}&1000\ \mbox{kg.m}^{-3} & & \\
    \hline
  \end{array}
  $$
  \caption{Summary of  model parameters.\label{tab:parametres}}
\end{table}

 The initial compositions of the droplet $\phi_{A}^{ini}$ and $\phi_{B}^{ini}$ must satisfy the following  conditions. No phase separation should occur during the whole simulation neither in the droplet nor in the bulk phase. Therefore, the initial compositions must be outside the spinodal region and the path between the initial and the equilibrium compositions must  not cross it. If ($\phi_{A}^{ini},\phi_{B}^{ini}$) is inside the spinodal region, the droplet will separate immediately. In addition, if $(\phi_{A}^{ini},\phi_{B}^{ini})$ is outside the miscibility gap but on the same side as the equilibrium point of the continuous phase, the system will evolve  towards one single-phase system. Consequently, to reproduce the desired dynamics, the initial compositions of the two phases must necessarily lie on both sides of the spinodal region. 

Taking these criteria into account, we have chosen a droplet with 85\% of element B and 15\% of element A. Doing so, the dispersed and continuous phases will reach their respective equilibrium composition following the arrows as shown in fig. \ref{fig:paysage thermodynamique0.15_0.85}. These are obtained from the orientation of the ${G}$ gradients at the initial composition points. This ensures that no water is introduced into the droplet. With this choice  the  initial density of the droplet is $\rho_{dis}^{ini}=980$kg/m$^{3}$ (slightly lower than that of water). 

\section{Simulation setup}

In this work, we use an inhouse pseudospectral code that was designed to solve the coupled binary Cahn-Hilliard Navier-Stokes model and has been extended to study ternary Cahn-Hilliard systems for the purpose of this study. \change{ The prediction of the code were also tested against the linear stability analysis of the Rayleigh Taylor instability for viscous fluids (at scales where diffusion can be neglected) and very good agreement was reached\cite{zanella2020,zanella2021}.  In addition as described in Supplemental Material [URL to be inserted by the editor] the code was shown to reproduce quantitatively well the initial stages of phase separation\cite{Cahn1961} in a ternary system (at scales where flow effect can be neglected when compared to diffusive mass transport).}

It is worth noting that the study of this configuration was initiated within the framework of Rasolofomanana's Ph.D. thesis \cite{rasolofomanana2023a} using the Cahn-Hilliard model \cite{rasolofomanana2022} of the TrioCFD code \cite{angeli2015}. This model was extended from binary case to n-ary case for this purpose. The numerical results presented here have been shown to be in good quantitative agreement with results of simulations with TrioCFD code. Only the pseudo-spectral code results are presented in this paper.


The computational domain consists of a 2D rectangular domain of width $L_x=54\times10^{-3}$ m and $L_y$, chosen large enough to avoid boundary effects in which a droplet of initial radius of $R=3.5 \times 10^{-3}$ m is placed. The wall boundary condition  are periodic at top and bottom. In order to avoid a global vertical motion of the fluid the zeroth mode of the buoyant term in the Fourier space is set to zero as in \cite{zanella2020}.  In addition a  no slip boundary condition at the lateral walls is imposed through  the introduction in the Navier-Stokes equation of a dissipative term $-\sigma v$ on a layer of 7 grid points with $\sigma dt=0.5$ which corresponds to an extremely fast  relaxation toward 0.

The viscosity of both phases is  $\eta=10^{-3}$ Pa.s. For such parameter values and a millimetric size droplet moving at velocities of the order of $10^{-2}$ m/s, the associated Reynolds number is of the order of $10$. This implies that inertial effects cannot be neglected. The surface tension is $\sigma = 36 \times 10^{-3}$ N/m as used in \cite{rao2015}. With this value and the properties of the droplet,  the Bond Bo and Morton Mo numbers are:
\begin{eqnarray}
Bo&=&\dfrac{\Delta \rho g D^2}{\sigma}\\
Mo&=&\dfrac{g\Delta \rho\eta^4}{(\rho_{cont}^2\sigma^3)}
\end{eqnarray}
where $\Delta\rho$ is  the density difference between two phases, $D$ is the diameter of the droplet and $\rho_{cont}$ the density of the continuous phase. In our case, we have $Bo\approx0.1$ and $Mo\approx 10^{-10}$ corresponding to the spherical regime according to Clift's diagram \cite{clift1978}. 

 The upscaling parameter $\lambda$ in  the phase field model allows to consider a  larger interface thickness than the characteristic length scale of the physical interface \cite{cardon2016,rasolofomanana2022a}. With such a thermodynamic landscape and the related equilibrium compositions, the parameters $\kappa^{bin}$ and $\lambda$ can be defined from the surface tension $\sigma$ and interface thickness $\varepsilon$ as:
\begin{equation}
\kappa^{bin} = \frac{3}{2}\sigma \varepsilon
\label{eq:kappa_potentiel_ternaire_}
\end{equation}
and 
\begin{equation}
\lambda=12\frac{\sigma}{\varepsilon}
\label{eq:parametre_upscaling_potentiel_ternaire_}
\end{equation}
The numerical results are expected to converge toward the actual solution when $\varepsilon\to0$ provided that the mobility $M$  is properly chosen, as will be discussed later.

The initial condition is a quiescent droplet with  composition:
\begin{widetext}
\begin{equation}
\phi_{i}(x,y,t=0)=\dfrac{\phi_{i}^{ini}}{2}\left(1-\tanh\left(\dfrac{\sqrt{(x-x_0)^{2}+(y-y_0)^{2}}-R}{\varepsilon}\right)\right)
\label{eq:goutte_tanh}
\end{equation} 
\end{widetext}
with $R$ the radius of the circular droplet, $\varepsilon$ the interface thickness, $(x_{0},y_{0})$ the coordinates of the center of the droplet in the 2D domain and  $\phi_{i}^{ini}$ is the composition of the element i$\in$\{A, B\} in the dispersed phase at initial state. Here $\phi_{A}^{ini}=0.15$  and  $\phi_{B}^{ini}=0.85$. 
With this choice of initial condition, assuming that the final droplet is composed of pure B \footnote{This is expected in an infinite domain with a finite size droplet and the thermodynamic landscape used here.}, neglecting the Gibbs Thomson effect, the final volume of the droplet is expected to be $(1-\phi_{A}^{ini})V_{dis}^{ini} $ with $ V_{dis}^{ini}$ the initial volume of the droplet. Finally  model  parameters  are summarized  in \Tab{parametres}.

The main quantities of interest in these simulations are the related to the droplet motion. Accordingly, since no mass transfer of the immiscible component is expected we  define the position of the droplet and its velocity by:
\begin{eqnarray}
   \bm{x_{G}}&=&\frac{\int \bm{x} \phi_B}{\int \phi_B}\\
    \bm{v_{G}}&=&\frac{\int \bm{u} \phi_B}{\int \phi_B}
\end{eqnarray}
where $\phi_B$ is the molar fraction of the immiscible component, $\bm{x}(x,y)$ and $\bm{u}$ are the spatial coordinates and velocities of a point. In the next section we discuss in more details the choice of parameters and the convergence of the model.

\section{Choice of parameters and convergence\label{secparamconv}}
\begin{figure}
  \centerline{\includegraphics[width=0.5\textwidth]{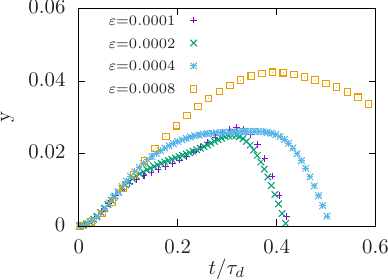}}
  \caption{position of the droplet center of mass along the $y$ axis as a function of $t/\tau_d=8.33$ (see eq.\ref{eq:taud}) for different values of the interface thickness $\varepsilon$.\label{fig:convergence}}
\end{figure}

  First we have checked the sensibility of our simulations to grid spacing and have found that using $dx= \varepsilon/4$ lead to converged results with respect to grid spacing: doubling the resolution did not change the results and the curves of the droplet position as a function of time  for both values of $dx$ superimposed. We have used this resolution throughout our simulations.  

Then we have checked the convergence of our simulations with respect to the interface thickness. To this purpose, we need to discuss the choice of the mobility $M$. Here the droplet motion evolution is due to diffusion coupled to Navier-Stokes equation. Using the  assumption that the interface is at equilibrium, the process that governs the motion is the diffusive mass transport far from the interface. In this region (far away from interface), the Laplacian term of the chemical potential can be neglected and the transport equation at a given point , using a second order Taylor  expansion of $G$ around the  concentration field at this point (with a diagonal $\mathcal{M}$) writes:
\begin{equation}
  \frac{\partial \phi_i}{\partial t}\approx M\Delta \lambda  \frac{dG}{d \phi_i} = (M\lambda) \left(\frac{d^2 G}{d \phi_i^2}\right)_{\phi_i} \Delta \phi_i=D_i\Delta \phi_i
\end{equation}
As a result, to keep the  bulk  diffusion kinetic  independent of the interface thickness $M\lambda \propto M/\varepsilon $ is kept constant when varying the interface thickness, keeping the surface tension $\sigma$ constant. With this choice of parameters we have checked the convergence of the simulations with respect to the interface thickness. To this purpose we have performed numerical simulations with the interface thickness ranging from 0.0001 to 0.0008 and, as can be seen in fig. \ref{fig:convergence} where the position of the droplet for $D_A=1.08. 10^{-6} m^2s^{-1}$ is plotted as a function of time. This value of $D_A$ corresponds to a diffusion time $\tau_d=8.33s$ defined as 
 \begin{equation}
   \tau_d=\frac{l^2}{D_A}\label{eq:taud}
 \end{equation}  
with $D_A$ the diffusion constant and $l$ a characteristic length taken equal to $10^{-3}m$ here. It is only for interface thickness lower than $2\, 10^{-4}m$ that the overall droplet trajectory is correctly captured. Comparing the droplet position obtained from simulations with $2\, 10^{-4}m$ and $1\, 10^{-4}m$, the droplet apex differs by $8\%$ and the associated time by $2\%$. For the purpose of this study, this level of accuracy is deemed acceptable and, in the following we present results that have been obtained using $\varepsilon=2\, 10^{-4}m$

It must be noted that the curve obtained for $\varepsilon=0.004$ is very similar (qualitatively and quantitively)  to the curve obtained for $\varepsilon=0.002$ and $\tau_d=11.11$ (not shown). Hence the lack of convergence could be related to an increased interfacial resistance when the interface thickness is increased  that is sufficiently high to slow down the diffusive transport. A similar behaviour is observed for heat transport  in the case of rough interfaces \cite{Kapitza}. 

 In the following we present results that have been obtained using $\varepsilon=2\, 10^{-4}m$.  While the situation discussed here cannot be described as at equilibrium  it is  worth mentioning that for this value, the  equilibrium  Gibbs concentration jump  at the interface is  of the order of $\approx 0.005$ for both A and B. For $\varepsilon=8\, 10^{-4}m$ the  concentration jump at the interface is   $\approx 0.01$. It  is proportional to the ratio between the interface thickness and the droplet   radius of curvature. 

 With this value of $\varepsilon=2\, 10^{-4}m$, $M$ is varied between $25\, 10^{-10}$ and $0.25\, 10^{-10}$, which corresponds to values of the diffusion constant $D_A= (M\lambda) \left({d^2 G}/{d \phi_i^2}\right)_{\phi_A=\phi_B=0} $ of the miscible component in pure  water  ranging from $1.08\,10^{-7}$ to  $1.08\, 10^{-5}m^2/s$ and Peclet numbers defined as $Pe=v_G R/D$, with $v_G\approx 10^{-2}m/s$ and $R\approx 3\, 10^{-3}m$, ranging from 3 to 300. With our choice of a symmetric $G$, the diffusion constant of component A in pure B is also $D_A$.  In the following, all results will be presented using $\tau_d$ as the control parameter for tuning diffusive transport.   In table \ref{table:mobs_diff}, we present the value of mobilities corresponding to a given diffusion constant $D_A$ and a given value of $\tau_d$ for two values of $\varepsilon$.

\begin{table}
	\begin{tabular}{|c|c|c|c|}
		\hline
		$M_{\varepsilon=0.0008}$ & 	$M_{\varepsilon=0.0002}$ & $D_A$ & $\tau_d$\\ \hline
		100.e-10     &   25.e-10 &      1.08e-5    &         0.833         \\ \hline
		50.e-10   &      12.5e-10    &  5.4e-6     &          1.666 \\ \hline
		20.e-10     &    5.e-10     &   2.16e-6     &         4.166\\ \hline
		10.e-10     &    2.5e-10   &    1.08e-6       &       8.333\\ \hline
		5.e-10     &     1.25e-10   &   5.4e-7         &      16.66\\ \hline
		2.e-10    &      0.5e-10  &     2.16e-7        &      41.66\\ \hline
		1.e-10     &     0.25e-10 &     1.08e-7          &    83.33\\	\hline
	\end{tabular}
	\caption{\label{table:mobs_diff} Table of the mobilities for two values of the interface thickness and the corresponding values of the diffusion constant $D_A$ and the corresponding diffusive characteristic time $\tau_d$. Units for diffusion constants are $m^2/s$, for $\tau_d$ they are seconds and mobilities are expressed in.}
\end{table}

\section{Results and discussion}
\label{sect:results}
   \begin{figure}
   	\centerline{
        \includegraphics[width=0.45\textwidth]{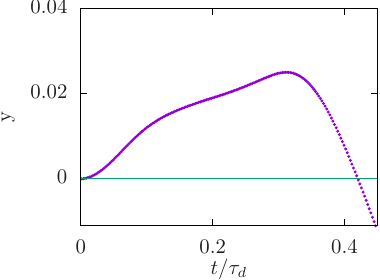}}
          \centerline{      \includegraphics[width=0.45\textwidth]{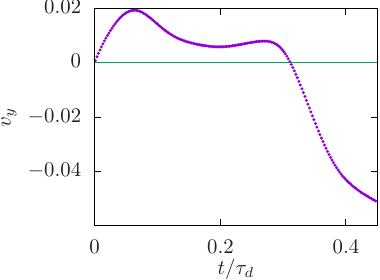} }
	\caption{Position (left) and speed (right) of the droplet as a function of time for $\varepsilon=0.0002$ and $\tau_d=8.333$. \label{fig:position}}
   \end{figure}
We now present numerical results obtained with this model and discuss them. 
\subsection{General behaviour} 
As expected, the position of the droplet center along the horizontal axis  remains constant throughout the simulation. In fig \ref{fig:position}  the position of the droplet along the vertical axis $y$ is plotted as a function of time for $\tau_d=8.33s$. Its trajectory includes an ascending phase followed by a descending phase. In the ascending phase, the droplet being lighter than the surrounding water first starts to rise. In the same time, the miscible component of the droplet diffuses out in water, resulting in an increase of the droplet density. Eventually, the droplet reaches a maximum height, becomes heavier than the continuous phase and ends up falling down into the water. This behaviour is in good qualitative agreement with experimental results presented in \cite{rao2015}. Due to the nature of our simulations and the value of parameters used here, quantitative agreement cannot be reached and this is out of the scope of this paper. Indeed, our simulations have been performed in 2D geometry  while experiments are 3D. In addition  the initial bubble composition used here differs significantly from the one in \cite{rao2015} and its initial mass density is 980kg.m$^{-3}$, which is heavier than in experiments. Some more information about the comparison between such Cahn-Hilliard based simulations (carried out with TrioCFD) and the experimental results can be found in \cite{rasolofomanana2023a}.
   \begin{figure*}
   	\centerline{
   		\includegraphics[width=.9\textwidth]{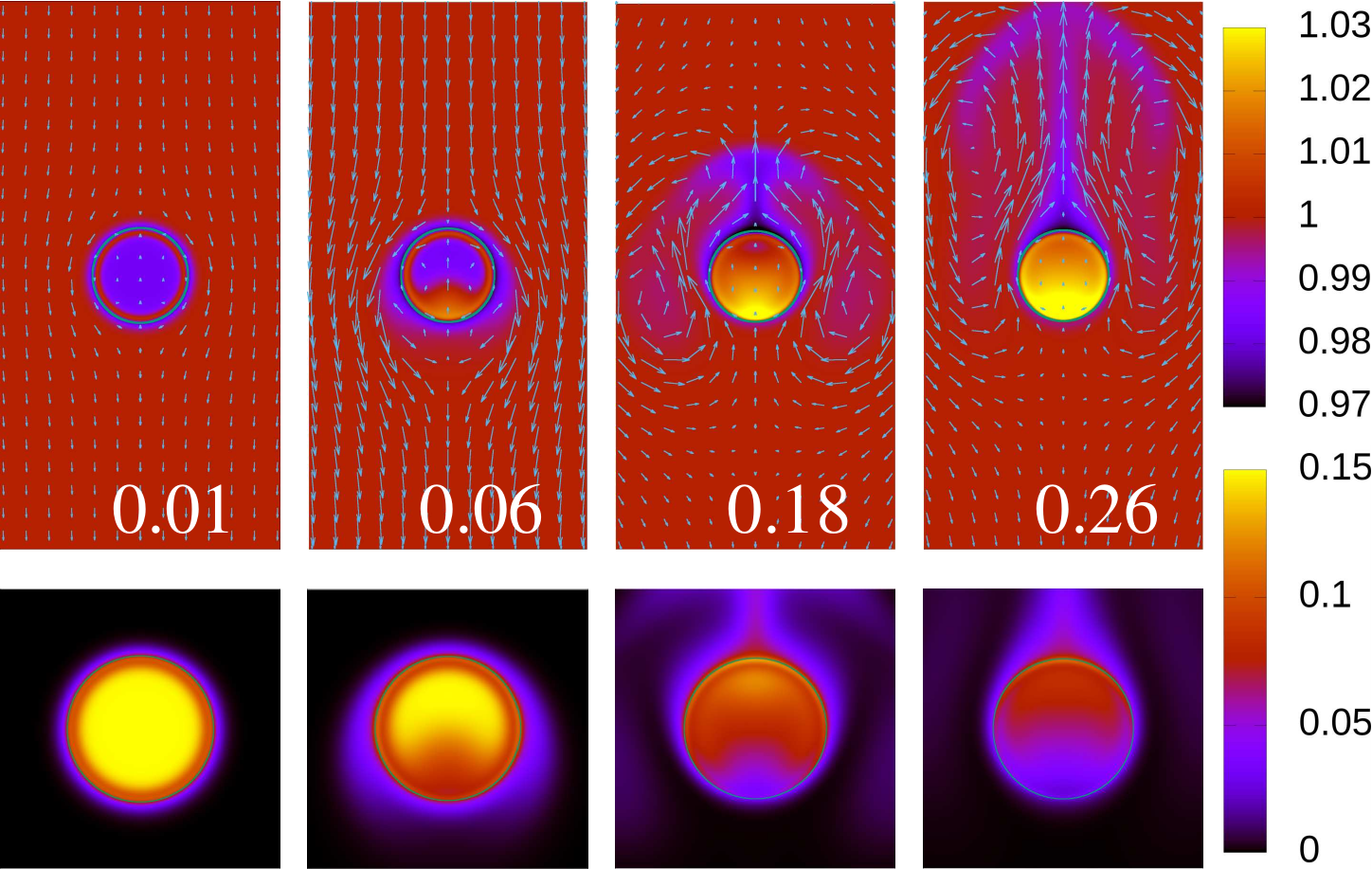} }
                \caption{Colormaps of the fluid density (top row) and of the A component concentration (bottom row) at different rescaled times $t/\tau_d$(the values label the density maps). The droplet  (the B rich region) is shown by the green circle.\label{fig:maps}. Parameters are the same as in figs. \ref{fig:position}. The structure of the flow is illustrated by the flow field with respect on the top row in the frame. The represented field is $\mathbf{u}-v_y\mathbf{e_y}$ where $v_y$ is the vertical velocity of the droplet.}
   \end{figure*}

We present a typical curve of the velocity along the vertical axis in fig. \ref{fig:position} for  $\tau_d=8.333$. After an acceleration phase, that is absent from experimental curves in \cite{rao2015}, the velocity of the droplet reaches a maximum and starts to decrease. However contrarily to experimental results the decrease of the velocity is not as regular as it would be expected from a na\"ive approach. Indeed, the evolution of the droplet velocity is clearly non monotonous and presents a bump before a fast acceleration of the droplets leads to its fall (it eventually reaches a terminal velocity). In order to better understand  this unexpected bump we present concentration maps of $A$ together with the flow field  and fluid density maps at different times of the evolution of the droplet in fig. \ref{fig:maps}. The colormaps are taken at the beginning of the motion, when the droplet velocity reaches its maximal value, after it has decreased and reaches a first minimum and close to the moment of the bump in the droplet velocity. Hence the discussion here is focused on the upward motion of the droplet.

The first observation is that the droplet presents in heterogeneous concentration in $A$ with a minimum close to its bottom. In addition, one can see that at the interface there is a maximum of A concentration that is similar to the bumps that can be seen in some concentration profiles at equilibrium \cite{rasolofomanana2022a}. Before going further we want to discuss the issue of these bumps.  Since  the thermodynamic landscape has been chosen to avoid such bumps they do not share the same cause  as the one observed in \cite{rasolofomanana2022a}. An analysis using the chemical potential of $A$ shows that the concentration bump do not correspond to any chemical potential bump and that when going through the interface, the gradient of chemical potential is constant which corresponds to a constant mass flux when crossing the interface as expected. Hence these bumps are simply related to the fact that as mentioned in \cite{plapp2011a} when going through an interface the quantity that is continuous is not the concentration of a solute  but its  chemical potential. It must be noted that outside of the interface, the concentration of $A$ and its chemical potential are proportional to each other. 

  The higher concentration of A at the top of the droplet is in correlation with the fact that the $A$ diffusion layer is larger at the top of the droplet than at its bottom. This asymmetry of the diffusion layer  is in contradiction with what is expected  when considering the forced movement of an object that is releasing some chemical in the surrounding fluid: in this case the upper part of the droplet is advancing in a fresh fluid and the diffusion layer ahead of the droplet should be thinner than behind it. Here the opposite is observed. This is due to the fact that the flow is affecting the diffusive layer. Indeed, since $A$ is lighter than water (and than the droplet), there is an upward motion of the diffusion layer with respect to the drop  that leads to the formation of the thicker layer of $A$ above. This illustrated by the flow pattern where one can see that close to the droplet the flow is moving upward with respect to the droplet due to buoyant forces.      
     
 These observations already illustrate the effect of  diffusion on flow. On the density map, one can also see effects of the flow on diffusion. Indeed,  the droplet density is asymmetric and that it is denser in its lower part. This is related to the fact that, as seen on the map, the diffusive layer is thinner at the bottom of the droplet than at its top. This leads to a more efficient diffusive transport of A out of the droplet, a faster decrease in A at the bottom of the droplet, which translates into a heavier fluid. This configuration is stable with respect to buoyant forces, however, its origin is  the interplay between diffusion and flow and not some sedimentation. It must be noted that the asymmetry of the diffusion layer is the opposite of what is expected if there is no effect of the concentrations on flow. One would expect a larger concentration of $A$ behind the droplet. 

\subsection{ Effect of the mobility}
     
\begin{figure}
	\centerline{
        \includegraphics[width=0.45\textwidth]{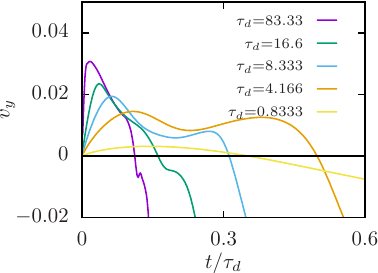}}\centerline{\includegraphics[width=0.45\textwidth]{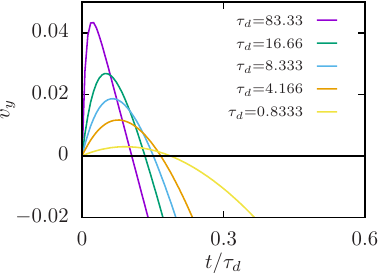} }
        \caption{\textbf{left} Absolute velocity of the droplet for different mobilities with respect to time $\times$ mobility. \textbf{right} velocity curves obtained using the reduced model.\label{fig:reduceddatas}}
\end{figure}

\begin{figure}
        \includegraphics[width=0.45\textwidth]{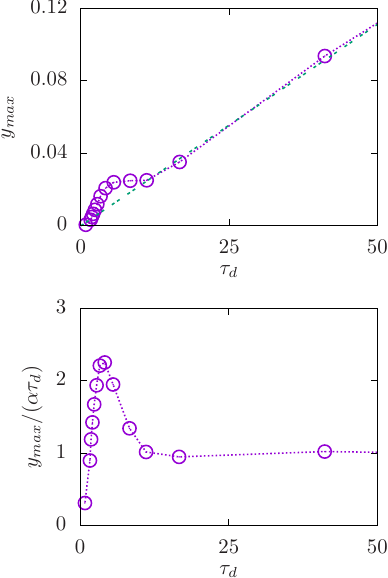} 
        \caption{\textbf{left} Position of the maximum $y$ of the trajectory of the droplet as a function of the diffusion time $\tau_{d}$. The line, of equation $y=\alpha \tau_d$ with $\alpha=0.002222$ corresponds to a linear fit of the data for the lowest mobilities. \textbf{right} Ratio of the maximum $y$ of the trajectory with the fit for low mobilities. \label{fig:ymax}}
\end{figure}

   In order to illustrate the effects of the interplay between flow and diffusion we find it interesting to first consider a system where such effects are absent. To this purpose we consider a simple model that describes the inertial motion of an object that is gaining mass along time and that is subject to  a viscous drag force. The equation of motion of this object writes:
\begin{equation}
  \dfrac{du}{dt}=f(t)- \frac{u}{t_r}
\label{eq:equation_masse_modele_reduit}
\end{equation}
where  $f(t)$ is a function that describes the effect of buoyant forces and that evolves due to diffusive process and $t_r$ is the characteristic time associated with the effective viscous drag. We consider the simple case where the driving force for the motion  $f(t)=\frac{v_0}{t_r}(1-t/t_0)$, with  $v_0$ a characteristic velocity. With this expression,  $t_0$ is a characteristic time that is proportional to $\tau_d$. Taking into account the initial condition $u(0)=0$, this simple model can be solved analytically and has for solution~:
\begin{equation}
u(t)=v_0 \left[\left(1+\dfrac{t_r}{ t_0}\right)\left(1-e^{- t/t_r}\right)-\dfrac{ t}{t_0}\right]
\label{eq:modele_reduit}
\end{equation}
We have chosen to determine   the model parameters  with a fit of the early stage  (i.e. the acceleration part and the onset of the slowdown) of simulations using  an intermediate value of  $\tau_d=8.333s$. A good fit is reached for the following parameter values: $t_0=0.806$, $v_0=0.056$ and $t_r=0.709$. Using these values and considering that   $t_0$ is proportional to the diffusion time, we plot the curves of the velocity as a function of time rescaled by $\tau_d$ for different values of $\tau_d$ on the left panel of  fig. \ref{fig:reduceddatas}. On the bottom panel the actual numerical results are plotted. There is  a significant discrepancy between the results of the reduced model and the actual numerical results. It is simply quantitative for high and low values of the characteristic time for  diffusion $\tau_d$. But for intermediate values the simple model does not capture the bump in the velocity curve that is seen and has already been discussed in the previous section. This bump has significant effects since it corresponds to a significant increase in the maximum height reached by the droplet. This can be seen  in fig. \ref{fig:ymax} where the maximum height is plotted as a function of the diffusion time.

For low mobilities (high $\tau_d$), the maximum height scales linearly with $\tau_d$, which is expected from a simple scaling argument since $\tau_r<<\tau_d$. For intermediate mobilities, the maximum height is significantly higher than this linear scaling (up to 2.5 times higher). While for low values of $\tau_d$  (smaller than $t_r$) the maximum height scales like $\tau_d^2$ due  to inertial effects: the times it takes for the bubble to accelerate is larger  than $\tau_d$. 

\begin{figure}
  \includegraphics[width=0.45\textwidth]{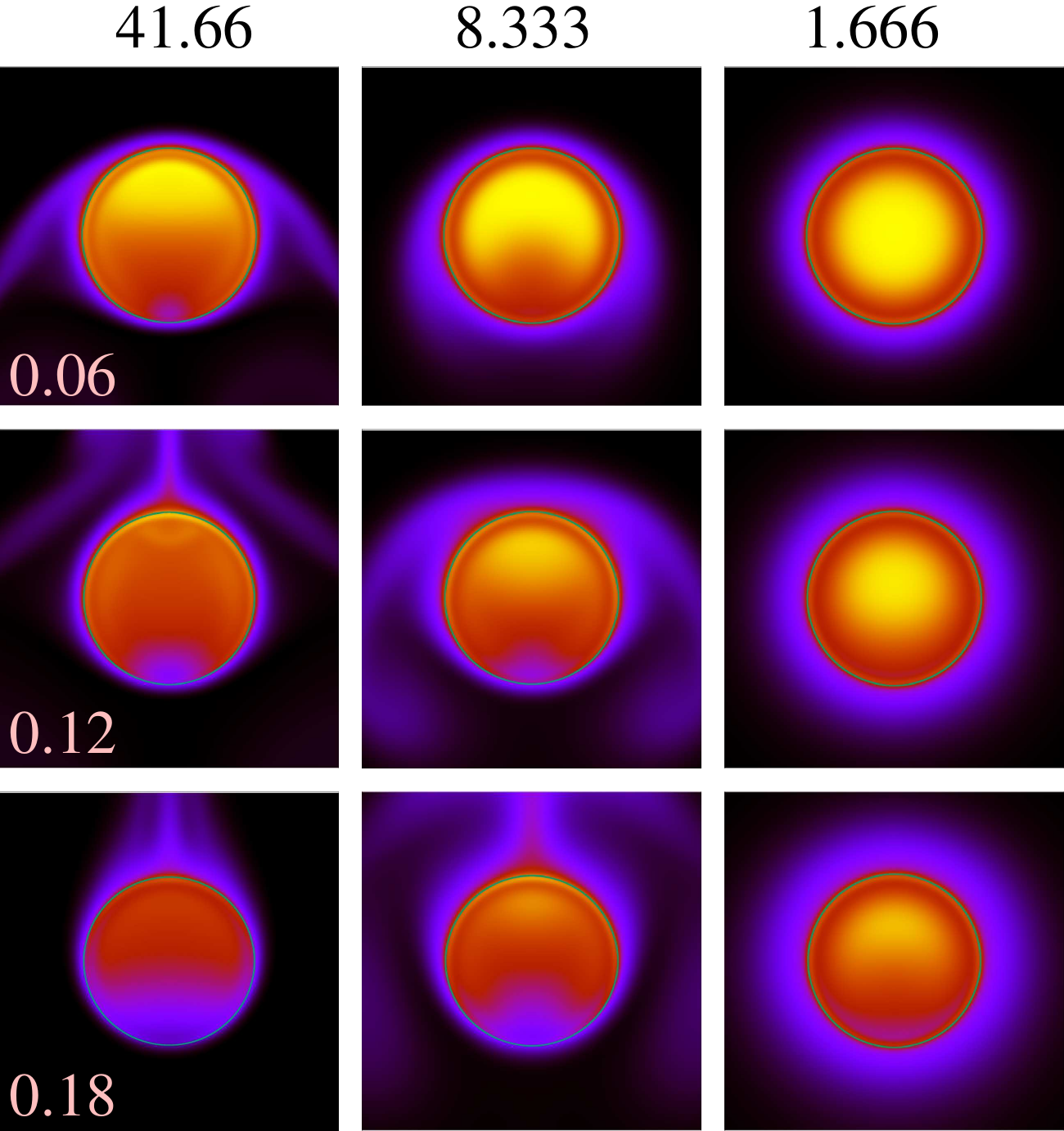} 
  \caption{colormaps of the A concentration for three different diffusion times  ($\tau_d=8.33,\ 41.66 \mbox{ and } 1.66$ ) at three different dimensionless times  $t/\tau_d=0.06,\ 0.12 \mbox{ and } 0.18$. The colorscale is the same as in fig. \ref{fig:maps}.} \label{fig:concA}
\end{figure}

Finally, as its importance on the motion of the droplet has been made obvious here, we find it useful to discuss the diffusion layer around the droplet for different values of the mobility. To this purpose, we have plotted the diffusion layer for $\tau_d=8.33,\ 41.66 \mbox{ and } 1.66 s$ at $t/\tau_d=0.06,\ 0.12 \mbox{ and } 0.18$ in fig. \ref{fig:concA}. From these maps, one can see first that for slow diffusion, the flow leads to the  thinning of the diffusion layer around the droplet (1st line). This in turn leads to higher gradients in the bulk and therefore to higher A flux outward of the droplet. As a result, as can be seen on the last line, for higher mobility the amount of A in the droplet at a given rescaled time is higher. When one consider the diffusion layer, it is also clear that it presents a vertical asymmetry: it is thicker above the droplet (which is moving upward) while in the case of a passive diffusing component it should be thicker below. This illustrate the fact that the lighter A rich fluid around the droplet  is moving upward. As a result the diffusive transport of $A$ is weaker in the region above the droplet and there is an higher concentration of  $A$ in the top part of the droplet.

\begin{figure}
  \includegraphics[width=0.45\textwidth]{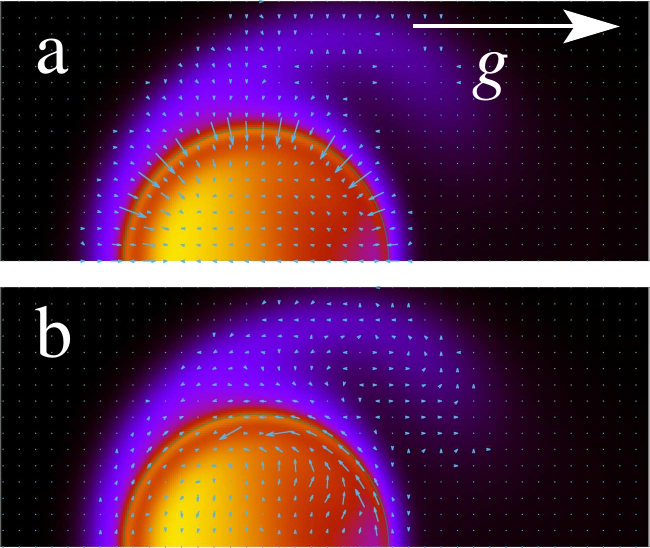} 
  \caption{\textbf{a} Colormaps of the A concentration together with the vector plot of source term for surface tension. \textbf{b} same plot  with the source term projected along the tangent to the interface (i.e. normal to $\nabla \phi_B $) and magnified by a factor of 25 when compared to \textbf{a}. The parameters are $\tau_d=8.333$ and $t/\tau_d=0.096$. At this time the droplet is moving upward. The orientation of the gravity field is given by the arrow.} \label{fig:marangoni}
\end{figure}

This translates into solutal Marangoni forces that  affect the motion of the fluid.  These forces are due to the $\Sigma \tilde{\mu}_i \nabla \phi_i$ term in eq.~\ref{eq:NS1} that is the term that leads to the surface tension. This vector  field is represented in fig. \ref{fig:marangoni} a together with the concentration map of the miscible component and one can see that it leads to mostly inwardly    pointing vectors located at the interface.   In order to keep the figure readable the vectors are plotted only one every 8 point in each direction. In fig. \ref{fig:marangoni} b,  to represent Marangoni forces these vectors are plotted after a projection on the direction tangential to the interface (the gradient of $\phi_B$ ) and a 25 times scaling. One can see that there is a component tangential to the interface that would correspond to a Marangoni force that tends to push the interface upward (which would reinforce the effect of the buoyant layer. However, as can be seen some of the  represented vectors show a discontinuity. This is due to the fact that for vanishing gradients of $\phi_B$ (in the vicinity of the interface) the projection operator  is discontinuous.  In addition with this approach the Marangoni forces are also present in the bulk and not located at the interface. Therefore while our analysis  indicates  that solutal  Marangoni forces are present and are contributing to the droplet motion, it is difficult to quantify their magnitude when compared to other sources of motion. In less complicated context, such as the bouncing droplet \cite{Lohse2}, we believe that our model could provide valuable insight. 


\section{Conclusion}

We have carried out a 2D simulation of the dynamics of a multicomponent droplet released in a water column using a ternary Cahn-Hilliard Navier-Stokes model. This model describes naturally coupled mass transfer and hydrodynamic phenomena.  To illustrate both the capacity of the model to solve such problem and the interest of fully coupled simulations we have considered the case of  a multicomponent droplet rising and sinking in water. We have first studied the model convergence with respect to the interface thickness and shown that for interface thickness less than $1/15$ the radius of the droplet, the results are well converged provided a proper scaling for the mobility is used. 

The numerical simulation results show that the  relative importance of  advection and  diffusion is a key parameter in the situation considered here. Accordingly, we have carried out a study with different mobilities in order to understand how the diffusion layer is affected by the flow. In our case of study where the flow is buoyancy driven, the effects of considering a fully coupled system go far beyond a correction of the parameters of a simplified model. Indeed, the coupling between the flow and the diffusion leads to a significant increase in the magnitude of the upward motion of the droplet before it starts to fall. The analysis of our results allows us to link this phenomenon with a drag effect of the light  diffusion layer around the droplet. 

Hence in this work we have shown that the use of fully coupled models for the evolution of multiphase flows can be necessary as soon as there is a significant diffusion from one phase to the other. This is  typically the case in many multicomponent systems. This work can find applications in many other fields of application ranging from the study of phase inversion of the corium pool and CO$_2$ sequestration in deep aquifers to microfluidic devices and many other \cite{Lohse1}.  In addition, the model studied here can be used in a 3D code without any change.

\begin{acknowledgments}
We would like to thank Elie Saikali from Universit\'e Paris-Saclay, CEA, Service de Génie Logiciel pour la Simulation, for his support in the development of the TrioCFD code for this work.
\end{acknowledgments}

\section*{Data Availability Statement}
Data   will be made available on https://entrepot.recherche.data.gouv.fr/dataverse/cnrs

\end{document}


\title{Supplementary material for:Rise and fall of a multicomponent droplet in a surrounding fluid: simulation study of a bumpy path}
\author{Mirantsoa Aimé Rasolofomanana}
\affiliation{CEA, DES, IRESNE, DTN, St Paul-Lez-Durance 13108, France}
\affiliation{Laboratoire de Physique de la Matière Condensée, Institut Polytechnique de Paris, CNRS, Palaiseau 91120, France}

\author{Romain Le Tellier}%
\affiliation{CEA, DES, IRESNE, DTN, St Paul-Lez-Durance 13108, France}


\author{Hervé Henry}
\affiliation{Laboratoire de Physique de la Matière Condensée, Institut Polytechnique de Paris, CNRS,  Palaiseau 91120, France}

\date{\today}

\maketitle
Here we first derive the growth rate for the ternary Cahn Hilliard diffusive equation used in the paper and compare it to numerical results  from the code used in order to produce numerical results. We show that the agreement is good. This indicates that the code used is working as it should for the diffusive process of a ternary system. 

\section{Linear stability analysis}
For the sake of simplicity we limit ourselves to the simple case used in the paper: both the gradient and  mobility matrices are  diagonal. In this case, the diffusive Cahn-Hilliard equation for a ternary system that writes:
\begin{eqnarray}
\frac{\partial \phi_{i}}{\partial{t}} &=& \nabla \cdot { M \nabla\tilde{\mu}_{i}} \nonumber \\ \label{eq:CH1} 
\tilde{\mu}_{i} &=& \frac{\partial {G}}{\partial \phi_i}-\kappa\Delta {\phi_i}
 \label{eq:CH2}
\end{eqnarray}	
where the total free energy of the system writes:
\begin{equation}
  \mathcal{F}=\int G(\phi_i)+\sum_i \frac{\kappa}{2}(\nabla\phi_i)^2 dx.
\end{equation}
With this evolution equation we consider the evolution of small perturbations of concentration $\delta \phi$ around an homogeneous state $\phi_0$. It obeys the linearised eq.\ref{eq:CH1} that writes
\begin{equation}
	\frac{\partial \delta \phi}{\partial{t}}=M \Delta H_{\phi_0} \delta \phi- M \kappa \Delta \left(\Delta \delta \phi \right)\label{eq:linstab}
\end{equation}
where $ H_{\phi_0} $ is the hessian matrix of $G$ in $\phi_0$. Since it is symmetric, it  has two real eigenvalues $\lambda_1$ and $\lambda_2$ and two eigenvectors $\Phi_1$ and $\Phi_2$. If one writes $\delta \phi$ in the base defined by $\Phi_1$ and $\Phi_2$: $\delta\phi=a_1\Phi_1+a_2\Phi_2$, the evolution equation of the Fourier transform of  $a_i$:$\hat{a_i}$ writes:
\begin{equation}
	\frac{d \hat{a_i}}{dt}=-M \lambda_i k^2 \hat{a_i} - M \kappa k^4 \hat{a_i}
\end{equation}
where $k$ is the wave number.
Considering the same $G$ as in the paper and $\phi_0=(0.5,0.1)$, one gets $\lambda_1=-0.48$ and $\lambda_2=-0.34$. As a result the maximum growth rate is supposed to be equal to 
\begin{equation}
s = M (-\lambda_1 k^2-\kappa k^4)
\end{equation}
\section{Numerical results}
\begin{figure}
  \centerline{
  \includegraphics[width=\textwidth]{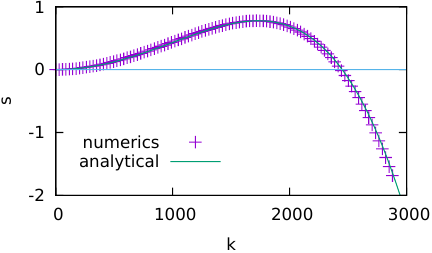}}
  \caption{\label{fig:comp}Comparison of the computed growth rate and the one predicted from the linear stability analysis. One can see a very good agreement.}
\end{figure}

Simulations were performed using as an initial condition a small perturbation (Gaussian noise with small amplitude) and the power spectrum (as a function of $||k||$) were computed at regular intervals. The growth rate for each value of $||k||$ was then computed using a linear fit in linear-logarithmic scale. In fig.\ref{fig:comp} the computed growth rate (divided by 2 to take into account that it corresponds to the growth of the squared amplitude) is plotted as a function of $k$ together with the theoretical growth rate. A very good agreement is reached. This indicates that the code used reproduces correctly the expected behaviour of the system as it should. The relative error is of the order of less than $5\%$. 

